%% file: main.tex
\PassOptionsToPackage{dvipsnames,table,xcdraw}{xcolor}
\documentclass[sigconf, screen, nonacm, 9pt]{acmart}
\AtBeginDocument{%
  \providecommand\BibTeX{{%
    Bib\TeX}}}
\setcopyright{acmlicensed}
\copyrightyear{2025}
\acmYear{2025}
\acmDOI{XXXXXXX.XXXXXXX}
\acmISBN{978-1-4503-XXXX-X/2018/06}

\input{preamble}

\usepackage{xcolor}
\usepackage{soul}
\usepackage{adjustbox}
\usepackage{bm}
\usepackage{algorithmicx}
\usepackage{algorithm}
\usepackage{algpseudocode}
\usepackage{pifont}
\usepackage{wasysym}
\usepackage{placeins} %
\usepackage{graphicx}
\usepackage{multirow}
\usepackage[hang,flushmargin]{footmisc}
\usepackage{balance}
\usepackage{latexsym}

\usepackage{multirow} %
\usepackage{url} %
\usepackage{subcaption}

\usepackage{tikz}

\usepackage{float}
\floatstyle{plain}
\newfloat{listing}{t}{lop}
\floatname{listing}{Listing}

\usepackage{listings}
\begin{document}
\raggedbottom

\newcommand{\abbr}{FLARE}
\title{FLARE: A Data\underline{F}low-Aware and Sca\underline{LA}ble Hardwa\underline{RE} Architecture for Neural-Hybrid Scientific Lossy Compression}

\author{Wenqi Jia}
\email{wenqi.jia@uta.edu}
\affiliation{%
  \institution{University of Texas at Arlington}
  \country{United States}
}

\author{Zhewen Hu}
\email{zhewen@tamu.edu}
\affiliation{%
  \institution{Texas A\&M University}
  \country{United States}
}

\author{Baixi Sun}
\email{sunbaix@iu.edu}
\affiliation{%
  \institution{Argonne National Laboratory}
  \country{United States}
}

\author{Yafan Huang}
\email{yafan@umd.edu}
\affiliation{%
  \institution{University of Maryland, College Park}
  \country{United States}
}

\author{Jiannan Tian}
\email{jtian@oakland.edu}
\affiliation{%
  \institution{Oakland University}
  \country{United States}
}

\author{Boyuan Zhang}
\email{boyuan.zhang@uky.edu}
\affiliation{%
  \institution{University of Kentucky}
  \country{United States}
}

\author{Daoce Wang}
\email{daocewang@unomaha.edu}
\affiliation{%
  \institution{University of Nebraska Omaha}
  \country{United States}
}

\author{Sian Jin}
\email{sian.jin@temple.edu}
\affiliation{%
  \institution{Temple University}
  \country{United States}
}

\author{Luanzheng Guo}
\email{lenny.guo@pnnl.gov}
\affiliation{%
  \institution{PNNL}
  \country{United States}
}

\author{Sheng Di}
\email{sdi1@anl.gov}
\affiliation{%
  \institution{Argonne National Laboratory}
  \country{United States}
}

\author{Yuede Ji}
\email{yuede.ji@uta.edu}
\affiliation{%
  \institution{University of Texas at Arlington}
  \country{United States}
}

\author{Miao Yin$^\dagger$}
\email{miao.yin@uta.edu}
\affiliation{%
  \institution{University of Texas at Arlington}
  \country{United States}
}
\renewcommand{\shortauthors}{Jia and Yin, et al.}

\begin{abstract}

Scientific simulations generate massive datasets whose storage and movement create major I/O and network bottlenecks. Neural-hybrid lossy compressors improve compression quality by combining conventional predictors with neural components, but their mixed control-flow, memory-bound, and neural workloads cause excessive data movement, pipeline stalls, and poor scalability on existing platforms. We propose \abbr, a dataflow-aware and scalable hardware architecture for neural-hybrid scientific lossy compression. \abbr~uses look-ahead computation ordering to reduce off-chip accesses, slice-wise normalization with operator fusion to remove global-normalization bubbles, and a modular hybrid dataflow to scale across data sizes and workloads. Across diverse datasets and platforms, \abbr~achieves $3.50\times$--$96.07\times$ runtime speedups and $24.51\times$--$520.68\times$ energy-efficiency improvements.

{\let\thefootnote\relax\footnotetext{$^\dagger$ Corresponding author.}}

\end{abstract}


\maketitle

\section{Introduction}

Scientific simulation powered by high-performance computing (HPC) enables accurate modeling of complex phenomena in climate science, physics, biology, and cosmology~\cite{cappello2025lossy, rosa2020data, di2024survey, almgren2013nyx, sexton2021nyx, lukic2015lyman, onorbe2019inhomogeneous}. However, modern simulations routinely generate petabyte-scale data, such as CMIP climate outputs~\cite{meehl2000coupled} and large fluid dynamics runs~\cite{johnson2023exploring}. Storing and moving these data stresses shared I/O and network resources~\cite{kim2023design, behzad2019optimizing, paul2020understanding, kang2022study, chunduri2019gpcnet, wang2025stz, zhang2026opal, yan2025eureka, yang2024seqbench}, making efficient data compression essential for scalable scientific workflows.

Lossless compressors~\cite{ziv1977universal, GZIP1996, lindstrom2006fast, gailly2004zlib} provide limited compression ratios on scientific data~\cite{zhao2020sdrbench}. Error-bounded lossy compressors~\cite{lindstrom2014fixed, di2016sz, liu2022qoz, zhao2021splinesz, tao2017significant, liang2019hybrid, tao2019optimizing, liang2022sz3, jiao2025adaptive, yang2025ipcomp, wu2025unstructured, zhang2025lcp, xia2024determinant, xia2025tspsz, xia2026trajectories} achieve higher ratios using local predictors such as curve fitting~\cite{di2016sz} and spline interpolation~\cite{lakshminarasimhan2013isabela}. Yet closed-form predictors can miss irregular features. Recent neural-hybrid methods therefore integrate DNN predictors~\cite{hornik1989multilayer, liu2021autoencoder, liu2023srnsz, li2024attention, gwlz, jia2024neurlz, liu2025crossfield} to improve compression quality while preserving error bounds.

Despite their benefits, neural-hybrid compressors run inefficiently on general-purpose platforms. CPU-only systems lack the throughput and parallelism needed for modern data rates~\cite{liu2024cusz, tian2020cusz, CPU-HBM-not-working}; GPU-only systems accelerate neural operators but underutilize resources on sequential and memory-bound non-neural stages; and CPU-GPU pipelines suffer from PCIe transfers between neural and non-neural components. These mismatches motivate customized hardware support~\cite{huang2024cuszp2}.

While an application-specific ASIC appears to be a natural solution to the limitations of general-purpose processors, a naive implementation of neural-hybrid lossy compression faces significant challenges. \ul{First}, the level-wise predictor introduces a memory-bound bottleneck due to irregular access patterns and extensive intermediate data that saturate on-chip buffers and exert excessive pressure on off-chip bandwidth. \ul{Second}, the absence of inter-stage pipelining leads to severe idle cycles, as each module -- predictor, neural network, and lossless codec -- must wait for its predecessor to finish. \ul{Third}, fixed-function datapaths hinder scalability, making it difficult to adapt to varying data sizes and workloads without costly hardware redesign.

\begin{figure}[t]
  \centering
  \includegraphics[width=\linewidth]{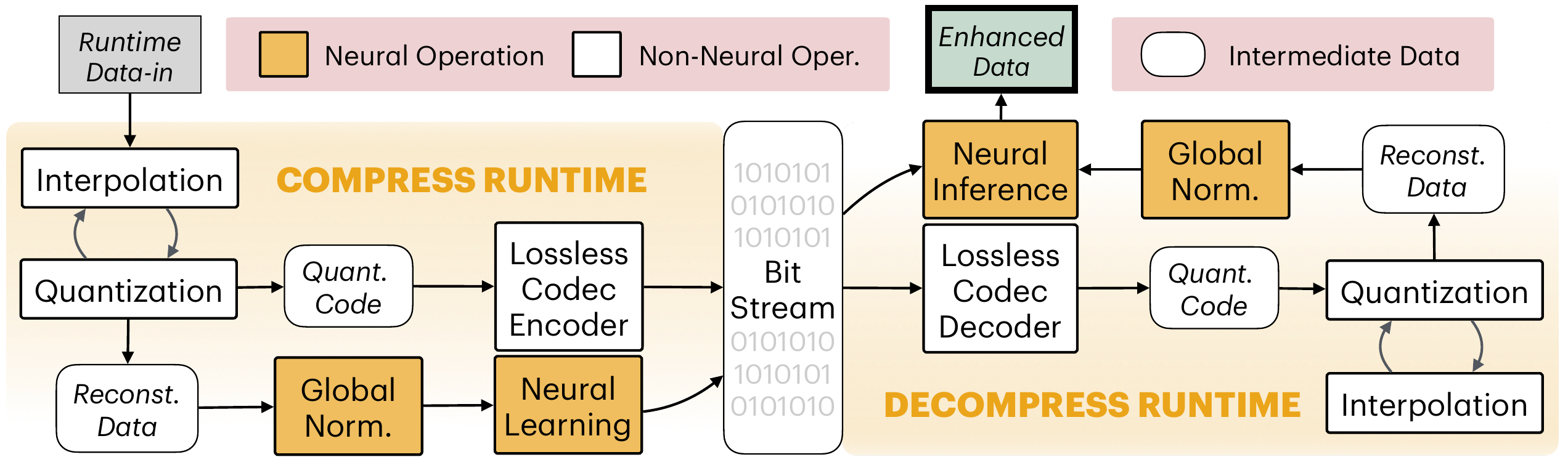}
  \vspace{-6mm}
  \caption{Overview of the state-of-the-art neural-hybrid lossy compression method, NeurLZ. "Global Norm." denotes the global normalization step.}
  \label{fig:neurlzOverview}
  \vspace{-3.5mm}
\end{figure}

To overcome these limitations, we propose \abbr, a dataflow-aware and scalable hardware architecture for neural-hybrid scientific lossy compression. \abbr~introduces \textbf{look-ahead computation ordering} to reduce intermediate buffering and off-chip DRAM accesses, \textbf{slice-wise normalization with operator fusion} to remove global-normalization stalls and redundant memory movement, and a \textbf{dataflow-aware scalable design} to coordinate prediction, neural processing, and lossless coding through hybrid pipelined-parallel execution.
To our knowledge, \abbr~is the first scalable hardware architecture for neural-hybrid scientific lossy compression, achieving $3.50\times$--$96.07\times$ runtime speedups and $24.51\times$--$520.68\times$ energy-efficiency gains.

\section{Background and Motivation}

\subsection{Neural-Hybrid Lossy Compression}

\begin{figure}[t]
  \centering
  \includegraphics[width=\linewidth]{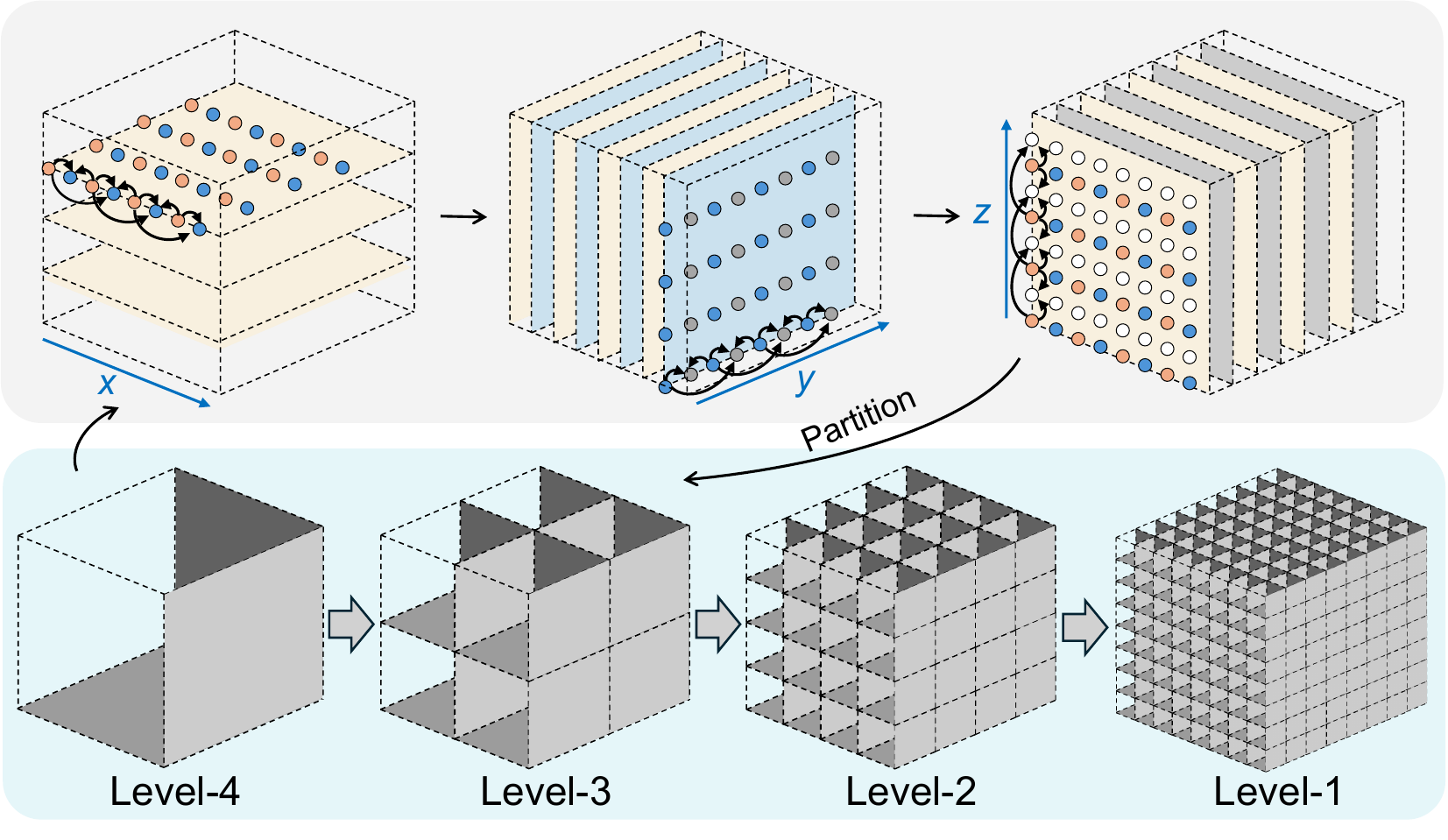}
  \vspace{-8mm}
  \caption{Level-wise, block-based interpolation. The upper part illustrates interpolation in a single block.}
  \vspace{-4mm}
  \label{fig:originalComputationOrder}
\end{figure}

The state-of-the-art neural-hybrid lossy compressor, NeurLZ~\cite{jia2024neurlz}, builds on SZ3~\cite{liang2022sz3} by combining interpolation-based prediction, quantization, lossless coding, and lightweight neural operators, as shown in Figure~\ref{fig:neurlzOverview}. During compression, NeurLZ performs level-wise, block-based iterative interpolation (Figure~\ref{fig:originalComputationOrder}); each prediction error is quantized, the quantization code is entropy-coded, and the reconstructed value is reused as a reference for later levels. After conventional reconstruction, the full data are globally normalized and processed by the neural module for online feature learning. Decompression reverses this flow: quantization codes are decoded, values are iteratively reconstructed, and the globally normalized reconstruction is refined by the stored neural model.

\subsection{Analysis of General Hardware}

\begin{figure}[t]
  \centering
  \includegraphics[width=\linewidth]{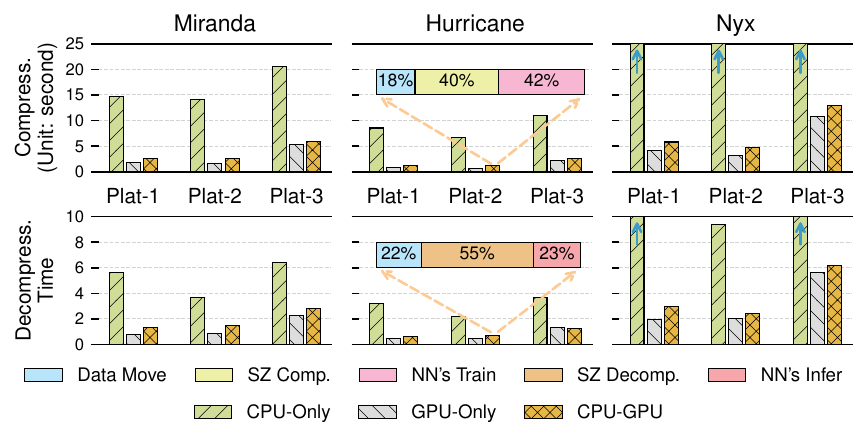}
  \vspace{-8mm}
  \caption{Processing time of three platforms across three common datasets on CPU-only, GPU-only, and CPU-GPU systems. The breakdown details the time composition of CPU-GPU on Hurricane at Plat-2.}
  \vspace{-6mm}
  \label{fig:timeComparison}
\end{figure}

General-purpose CPUs and GPUs poorly match NeurLZ's hybrid workload, which mixes control- and memory-bound non-neural stages with parallel neural computation. We evaluate three configurations on the platforms in Table~\ref{tab:platforms}: \textbf{CPU-only}, which preserves compression quality but lacks neural parallelism; \textbf{GPU-only}, which uses cuSZ~\cite{tian2020cusz} for non-neural stages and gives the shortest runtime but lower compression ratios; and \textbf{CPU-GPU}, which keeps high compression quality but suffers from PCIe transfers. On Hurricane at Plat-2, PCIe communication takes 18\% and 22\% of compression and decompression time, respectively. The compression ratios of CPU-only, GPU-only, and CPU-GPU are 387$\times$/25$\times$/387$\times$ on Miranda, 16,000$\times$/31$\times$/16,000$\times$ on Nyx, and 20$\times$/16$\times$/20$\times$ on Hurricane; runtimes are shown in Figure~\ref{fig:timeComparison}.

\begin{small}
\begin{table}[b]
  \vspace{-2mm}
  \caption{CPU and GPU Specifications for Evaluation.}
  \vspace{-0.5\baselineskip}
  \label{tab:platforms}
  \renewcommand{\arraystretch}{1.2}
  \begin{tabular}{@{}lcc@{}}
    \toprule
        \textbf{Platform}& 
        \textbf{CPU Model}& 
        \textbf{GPU Model}\\
    \midrule
    Plat-1 & AMD EPYC 9254 & NVIDIA RTX 6000 Ada \\
    Plat-2 & Intel Xeon w5-3435X & NVIDIA RTX 4090 \\
    Plat-3 & Intel Core i7-5930K & NVIDIA GTX 1080 Ti \\
    \bottomrule
  \end{tabular}
\end{table}
\end{small}

\subsection{Challenges for Customized Hardware}

Customized hardware can exploit NeurLZ's structure, but a naive design faces three challenges.

\Circled{1} The workload is fundamentally \textbf{memory bounded}: level-wise block interpolation repeatedly reads and writes reconstructed data across levels, multidimensional interpolation creates irregular strides in linear memory, and global normalization adds two full-data sweeps. As shown in Figure~\ref{fig:timeComparison}, non-neural components in Hurricane decompression require only $0.23\%$ of the neural FLOPs but still consume nearly 2$\times$ latency, indicating that data movement dominates. Buffering all intermediate data on chip would require impractical SRAM capacity for large scientific datasets.

\begin{figure}[t]
  \centering
  \includegraphics[width=\linewidth]{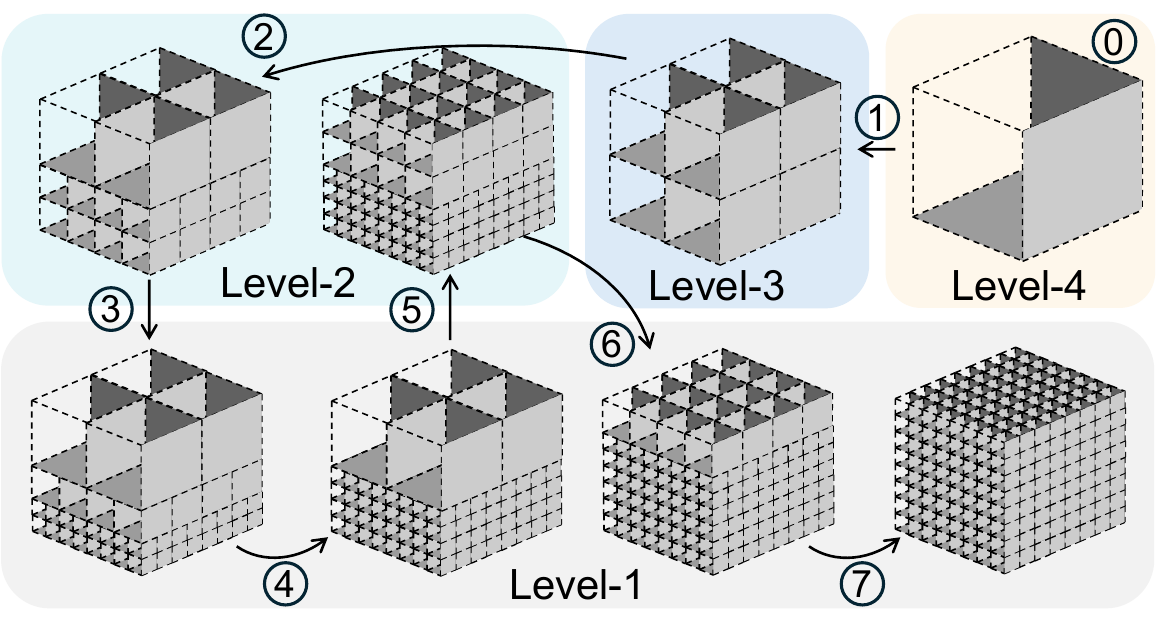}
  \vspace{-9mm}
  \caption{Proposed look-ahead computation order. Circled numbers indicate the execution steps.}
  \label{fig:proposedComputationOrder}
  \vspace{-1\baselineskip}
\end{figure}

\Circled{2} \textbf{Bubble overhead} arises due to the global normalization process, which prevents the system from maintaining a fully pipelined execution. Although partial quantized errors can be forwarded to the lossless encoder for efficient pipelining, global normalization requires access to the entire reconstructed dataset. This dependency delays the neural network, which processes data at a fine-grain two-dimensional slice level, until normalization is finished, thereby substantially reducing pipeline efficiency.

\Circled{3} \textbf{Limited scalability} in fixed-function datapaths hinders efficient handling of diverse data sizes and workloads. Over-provisioned designs lead to unnecessary silicon area and energy overhead, while under-provisioned ones fail to meet latency constraints.

These challenges underscore the need for a more efficient, scalable, and dataflow-aware architecture to support neural-enhanced lossy compression at scale.

\section{Proposed Hardware: FLARE}

\subsection{Look-ahead Computation Order}

\begin{figure*}[t!]
  \centering
  \includegraphics[width=0.85\linewidth]{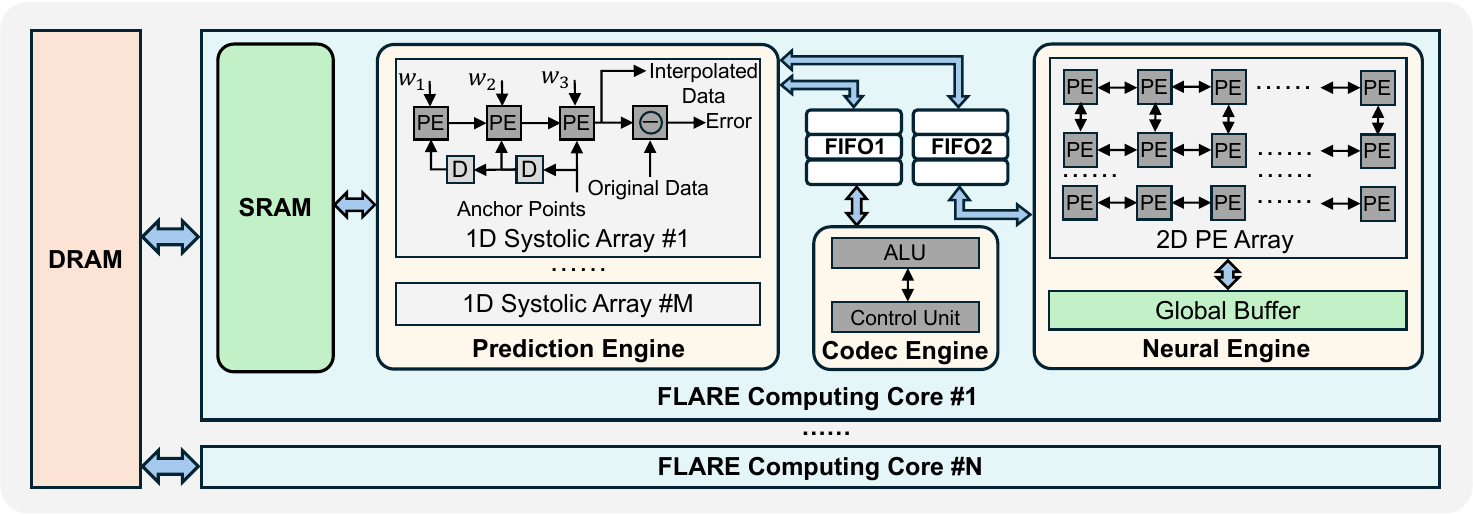}
  \vspace{-4mm}
  \caption{Architecture overview of the proposed FLARE.}
  \vspace{-4mm}
  \label{fig:architecture}
\end{figure*}

\begin{figure}[t]
  \centering
  \includegraphics[width=\linewidth]{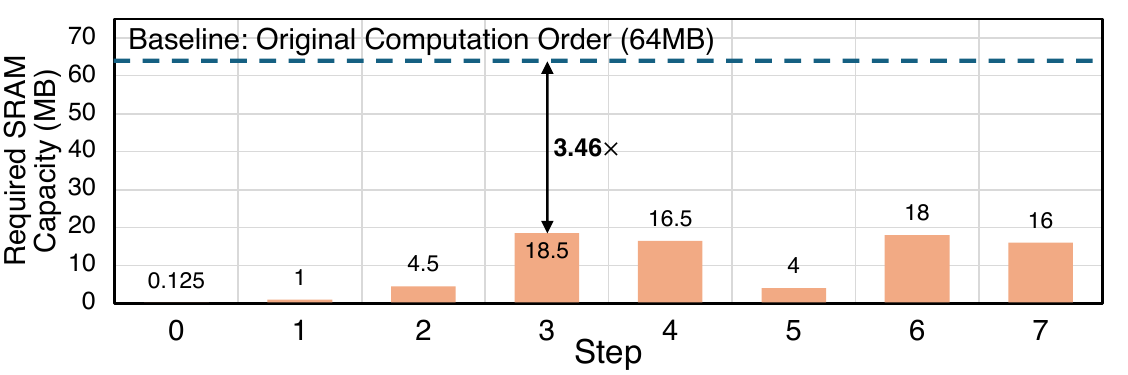}
  \vspace{-8mm}
  \caption{Reduction in SRAM capacity by the proposed computation order compared to the baseline.}
  \label{fig:sramRequirment}
  \vspace{-4mm}
\end{figure}

\begin{figure}[t]
  \centering
  \includegraphics[width=\linewidth]{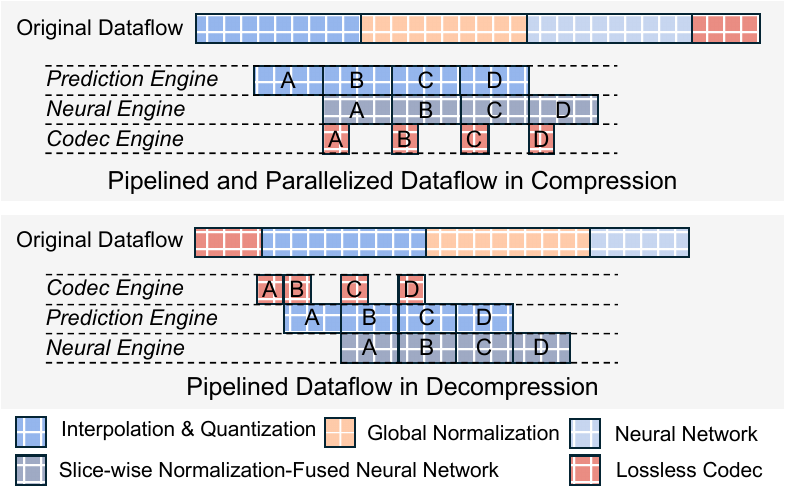}
  \vspace{-8mm}
  \caption{Compression and decompression dataflow.}
  \vspace{-4mm}
  \label{fig:dataflowExplaination}
\end{figure}

\begin{figure*}[t]
  \vspace{-4mm}
  \centering
  \includegraphics[width=\linewidth]{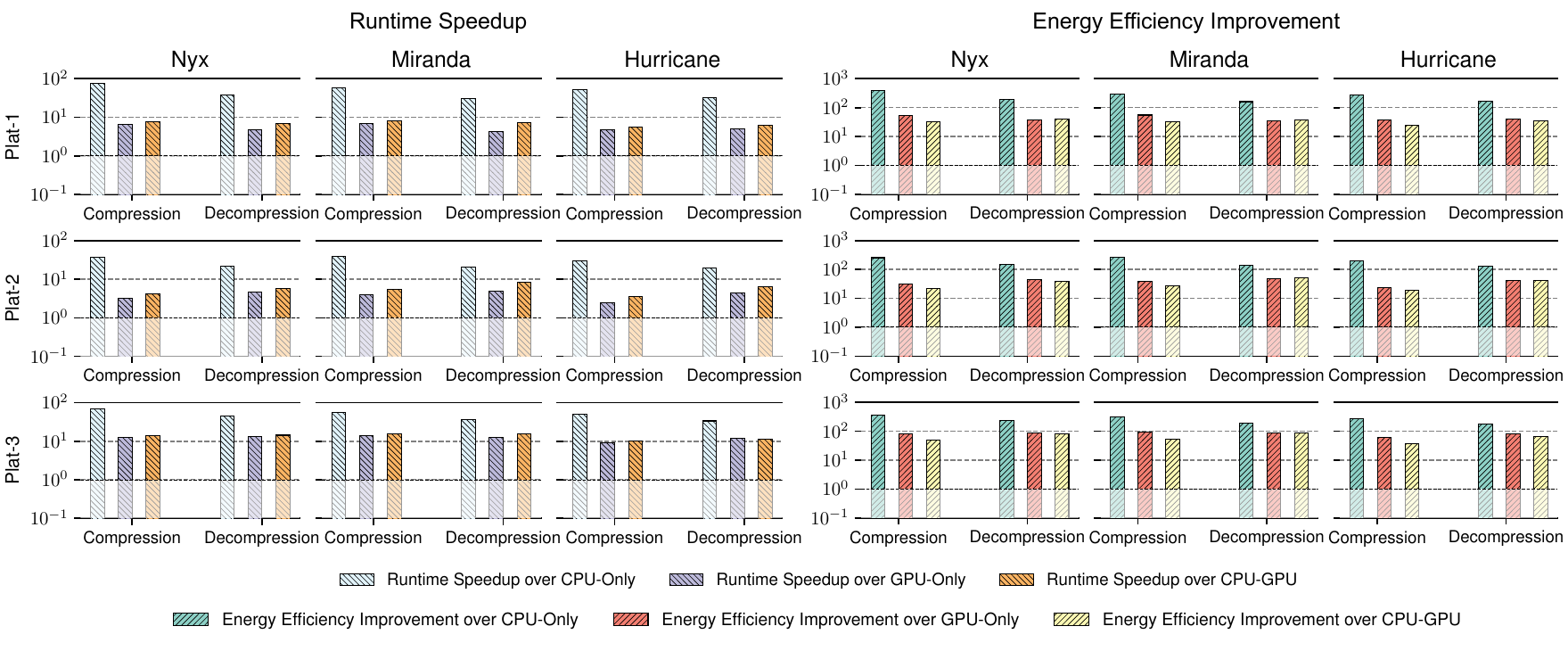}
  \vspace{-8mm}
  \caption{Evaluation of runtime speedup and energy efficiency.}
  \label{fig:efficiency}
  \vspace{-3mm}
\end{figure*}

\begin{figure*}[t]
  \centering
  \includegraphics[width=\linewidth]{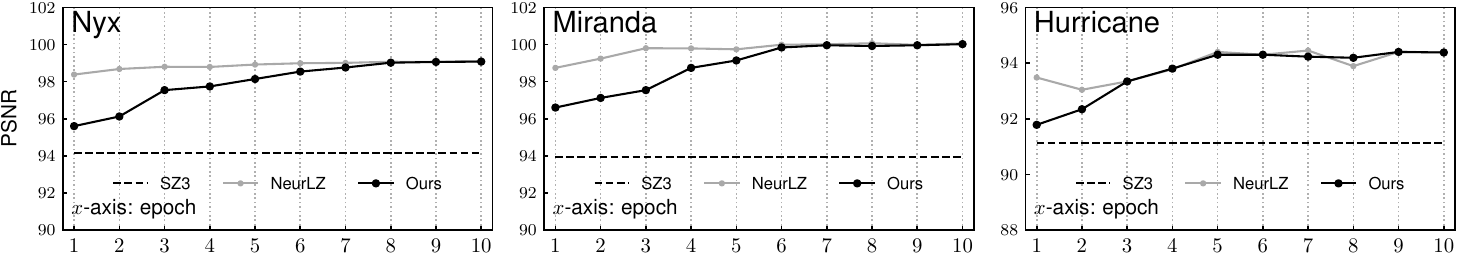}
  \vspace{-7mm}
  \caption{Algorithm performance curves of our proposed \abbr.}
  \vspace{-4mm}
  \label{fig:algorithmEvaluation}
\end{figure*}

To address intensive and irregular memory accesses under limited on-chip SRAM, we propose a look-ahead computation order that reduces the SRAM required for intermediate results while preserving algorithmic efficiency. The key idea is to exploit the data locality of the level-wise block-based interpolation procedure by reordering execution to maximize the reuse of reconstructed data. Since interpolation operations are independent across blocks, partial block-granularity computation can span multiple levels, allowing later levels to start before the current level is fully completed. As a result, the smaller intermediate results of partially processed blocks can be buffered in SRAM and reused early, avoiding expensive off-chip memory transactions. Once locality at Level-1 has been fully exploited, these partial results are directly forwarded to downstream modules because they are no longer needed for subsequent interpolation.

Figure~\ref{fig:proposedComputationOrder} illustrates the proposed look-ahead strategy. At each level, half of the newly reconstructed blocks are immediately forwarded to the next level, while the remaining half is deferred until the first half completes processing at Level-1 and produces partial final results. For example, the lower half of the blocks reconstructed at Level-3 is first processed at Level-2 (step 2), whereas the upper half is processed later at step 5 after the lower half has been computed through Level-1. Because Level-1 results are not needed for subsequent interpolation, this approach reduces the SRAM capacity requirement by 3.46$\times$ for a fixed block size of $32 \times 32 \times 32$ with 32-bit values, as shown in Figure~\ref{fig:sramRequirment}. In contrast, the original computation order requires the entire dataset to be stored in SRAM to avoid excessive off-chip memory accesses, which becomes impractical for large-scale scientific datasets.

Essentially, the proposed look-ahead computation order adopts a Depth-First Exploration strategy -- a recursive traversal that processes data blocks in a depth-first manner, producing partial results early in the computation. Unlike the original Breadth-First Exploration approach, which fully processes all blocks at each level before proceeding to the next, our depth-first approach enables an efficient execution order that exploits data locality to eliminate redundant data accesses, minimize memory usage, and enhance overall performance. By processing blocks across levels in an out-of-order fashion, the proposed algorithm effectively manages memory resources and reduces latency, making it particularly suitable for high-performance computing applications.

\subsection{Slice-wise Normalization with Operator Fusion}
\label{sec:operator_fusion}

Global normalization in the neural component scales and centers the data before neural network processing, but it requires repeated full-dataset traversals, causing redundant memory accesses and pipeline bubbles. To address this, we propose slice-wise instance normalization, which operates directly on two-dimensional data slices. This approach matches the neural network's slice-based execution pattern and enables fine-grained pipelined dataflow. Each normalized slice can then be streamed directly to the neural component on the fly, eliminating the overhead of global normalization across the entire dataset.

We further reduce memory accesses by fusing normalization with the first convolution kernel. Specifically, we use $max_i$ and $min_i$ to track the maximum and minimum values of the $i$-th slice during prediction. Embedding normalization into the convolution eliminates the separate read and write operations for normalized data. For each data point $D[x,y,i]$ in the $i$-th slice, the normalized value is computed as:
\begin{equation} 
  D_N[x,y,i] = \frac{D[x,y,i] - min_i}{max_i - min_i} 
  \label{equ:norm} 
\end{equation} 
where $D_N[x,y,i]$ denotes the normalized data. Integrating this normalization into the computation of the first convolution layer yields the convolution output $O[x, y, o]$:
\begin{align} 
  &O[x, y, o]  \notag \\
  &= \sum_{k_x,k_y} D_N[x + k_x, y + k_y, i] \cdot W[k_x, k_y, o] + b[o] \\ 
  &= \sum_{k_x,k_y} \frac{D[x + k_x, y + k_y, i] - min_i}{max_i - min_i} \cdot W[k_x, k_y, o] + b[o] \\
  &= \sum_{k_x,k_y} D[x + k_x, y + k_y, i] \cdot W'[k_x, k_y, o] + b'[o] 
\end{align} 
where the updated convolution weight $W'[k_x, k_y, o]$ and bias $b'[o]$ incorporate the normalization parameters and are computed as:
\begin{align}
  &W'[k_x, k_y, o] = \frac{W[k_x, k_y, o]}{max_i - min_i} \\ 
  &b'[o] = b[o] - \sum_{k_x,k_y} \frac{min_i}{max_i - min_i} \cdot W[k_x, k_y, o] 
\end{align}

Compared to explicitly computing $D_N[x, y, i]$, updating the convolution kernel $W'[k_x, k_y, o]$ and $b'[o]$ introduces minimal additional workload due to the small size. Furthermore, the compression quality achieved by the proposed instance normalization remains comparable to that of the original global normalization, as analyzed in Section~\ref{sec:algorithmEvaluation}.

\subsection{Dataflow-Aware and Scalable Design}

The proposed look-ahead computation order and slice-wise normalization allow reconstructed data normalization and quantized error generation to proceed at slice granularity. This enables the prediction modules (interpolation and quantization), lossless codec module, and neural network components in NeurLZ to execute under a hybrid dataflow that combines pipelined and parallel processing. During compression, the prediction modules generate a batch of slices at each time step, with the batch size determined by the block size. While subsequent slices are being computed, the current batch is dispatched downstream to the neural network and lossless codec module, forming a \textbf{pipelined dataflow}. At the same time, the neural network uses the reconstructed slices for error feature extraction, while the lossless codec, such as an entropy encoder, processes the quantized errors. This concurrent execution forms a \textbf{parallelized dataflow}. During decompression, the computation order is reversed: the dependency from Huffman decoding to data prediction creates a fine-grained \textbf{pipelined dataflow} across lossless decoding, prediction, and neural network inference. This hybrid execution pattern enables fine-grained processing in both compression and decompression, improving throughput and scalability.

To support this hybrid dataflow across diverse datasets and workloads, we propose a dataflow-aware, scalable \textbf{\abbr~Computing Core}, as shown in Figure~\ref{fig:architecture}. The core integrates prediction, including interpolation and quantization, with neural network and lossless codec components. By keeping intermediate results on-chip, each component can operate on the fly, reducing costly off-chip data movement and improving throughput. Although each module is customized for its own computation and memory access pattern, the overall architecture remains modular and scalable across different data sizes and workload configurations.

The \abbr~Computing Core consists of four components: (1) an on-chip \textit{SRAM buffer}, (2) a \textit{Prediction Engine} for interpolation and quantization, (3) a \textit{Neural Engine} for neural network operations, and (4) a \textit{Codec Engine} for lossless Huffman encoding and decoding.

The on-chip \textit{SRAM buffer} provides high-bandwidth, low-latency temporary storage for intermediate data. It reduces off-chip DRAM accesses and supplies fast data access to the Prediction Engine for pipelined execution.

The \textit{Prediction Engine} contains $M$ one-dimensional (1D) systolic arrays, each processing interpolation and quantization for one data block. Each array performs matrix-vector multiplications with fixed interpolation coefficients, such as $(w_1, w_2, w_3)$ in cubic interpolation. A sliding-window mechanism streams neighboring anchor points through the processing elements (PEs), improving data reuse. Within each array, the interpolation error is computed by comparing the interpolated values with the original data and is then quantized for data reconstruction. Since blocks are independent, the engine can process up to $M$ blocks in parallel.

The \textit{Neural Engine} accelerates neural network operations using a two-dimensional (2D) PE array for workloads such as convolutions, matrix multiplications, and attention mechanisms. Its grid layout supports parallel processing across multiple data streams. A dedicated global buffer caches intermediate results locally to reduce DRAM traffic and sustain throughput during training or inference.

The \textit{Codec Engine} uses ALU-based processing elements for control-intensive Huffman encoding and decoding. It is tightly integrated with the Prediction Engine so that quantized error values can be compressed or decompressed immediately. 

As shown in Figure~\ref{fig:dataflowExplaination}, during compression, outputs from the Prediction Engine are streamed slice-wise to both the Neural and Codec Engines for concurrent execution. Due to the look-ahead computation order, the Prediction Engine produces outputs at non-uniform intervals. To absorb this variability, two FIFO-based circular buffers, \textit{FIFO1} and \textit{FIFO2}, temporarily store quantized errors and reconstructed data, respectively, maintaining the pipelined dataflow. A similar pipeline is applied during decompression.

The \abbr~Computing Core scales along two dimensions. \textbf{Data-size scalability} is achieved by adjusting $M$, the number of parallel systolic arrays in the Prediction Engine, to match the throughput of the Neural Engine for different input sizes. \textbf{Workload scalability} is achieved by scaling $N$, the number of \abbr~Computing Cores, so multiple datasets can be processed concurrently. Thus, the architecture can adapt to different data sizes and workload intensities without redesign.

\section{Evaluation}
\subsection{Algorithm Performance Evaluation}
\label{sec:algorithmEvaluation}
We evaluate slice-based normalization in the NeurLZ framework by replacing global normalization with the method in Section~\ref{sec:operator_fusion}. Experiments use Nyx, Miranda, and Hurricane (Table~\ref{tab:dataset}) with a uniform error bound of $10^{-3}$ for SZ3, NeurLZ, and \abbr. Reconstruction quality is measured by PSNR.

\begin{small}
\begin{table}[t]
\caption{Property of tested datasets.}
\vspace{-0.5\baselineskip}
\centering
  \begin{tabular}{@{}lccc@{}}
  \toprule
  \textbf{Dataset}   & \textbf{Size (GB)} & \textbf{Dimension}          & \textbf{Domain}           \\ \midrule
  Nyx       & 0.5      & (512, 512, 512)    & Cosmology        \\
  Miranda   & 0.28     & (256, 384, 384)    & Large Turbulence \\
  Hurricane & 0.1      & (100, 500, 500)    & Weather          \\ \bottomrule
  \end{tabular}
  \label{tab:dataset}
  \end{table}
\end{small}

As shown in Figure~\ref{fig:algorithmEvaluation}, slice-based normalization improves PSNR over SZ3 by 6--10 dB across datasets. Its initial PSNR is slightly below NeurLZ, but becomes comparable after 5--6 epochs, showing that localized normalization preserves compression quality.
\subsection{Evaluation Setup for Architecture}
\textbf{Hardware configuration.}
We evaluate a single-core configuration ($N = 1$) with 32 MB SRAM, four 1D systolic arrays ($M=4$), 8 MB/32 MB FIFO buffers, and a $128 \times 128$ Neural Engine with a 24 MB global buffer. Table~\ref{tab:config} summarizes the parameters.

\begin{small}
\begin{table}[t]
  \centering
  \caption{Configuration of \abbr~Computing Core}
  \vspace{-0.5\baselineskip}
  \label{tab:config}
  \begin{tabular}{@{}llc@{}}
  \toprule
  \textbf{Component} & \textbf{Parameter} & \textbf{Value} \\ \midrule
  \multirow{1}{*}{SRAM} & Capacity & 32 MB \\
  \multirow{1}{*}{Interpolation Engine} & $M$ & 4 \\
  \multirow{1}{*}{FIFO1} & Capacity & 8 MB \\
  \multirow{1}{*}{FIFO2} & Capacity & 32 MB \\
  \multirow{2}{*}{Neural Engine} & Size of 2D PE Array & $128 \times 128$ \\
 & Global Buffer Size & 24 MB \\ \bottomrule
  \end{tabular}
  \end{table}
\end{small}

\textbf{Simulation and CAD tools.}
We build a bit-accurate, cycle-accurate simulator and a verified Verilog RTL implementation. The RTL is synthesized with Synopsys Design Compiler in a 28 nm CMOS library at 1 GHz, placed and routed with Synopsys IC Compiler, and evaluated for dynamic/static power using PrimeTime PX with simulation-derived switching activity. Memory area and power are estimated with Cacti, and neural dataflow is optimized with TimeLoop~\cite{parashar2019timeloop}. The final design reports \textul{7.38 W} power and \textul{100.07 mm\textsuperscript{2}} area.

\textbf{Baseline and benchmarks.} 
We use the same datasets and error bound as NeurLZ. Baselines include CPU-only, GPU-only, and CPU-GPU NeurLZ on the platforms in Table~\ref{tab:platforms}; the GPU-only baseline uses cuSZ~\cite{tian2020cusz}.

\subsection{Evaluation Result}

\textbf{Compression performance.}
We evaluate the compression performance of the proposed \abbr~across all systems and platforms using the Nyx, Miranda, and Hurricane datasets.  Although our method requires additional epochs to achieve the same PSNR as NeurLZ, it still delivers significant gains in both speed and energy, due to its optimized dataflow and architectural support. As shown in Figure~\ref{fig:efficiency}, on Nyx we observe $4.61\times$--$96.07\times$ runtime reduction and $31.27\times$--$520.68\times$ energy efficiency improvement; on Miranda, $5.75\times$--$81.48\times$ runtime reduction and $33.54\times$--$441.65\times$ energy efficiency gains; and on Hurricane, $3.50\times$--$68.16\times$ runtime reduction alongside $24.51\times$--$369.45\times$ energy efficiency improvements. These results confirm that our design not only accelerates lossy compression but also substantially lowers energy consumption.

\textbf{Decompression performance.}
\abbr~also achieves promising results in decompression. Specifically, on Nyx dataset, it delivers $4.69\times$--$37.96\times$ runtime speedups and $38.11\times$--$247.47\times$ energy efficiency improvements; on Miranda, $4.36\times$--$35.59\times$ runtime speedups and $35.48\times$--$192.89\times$ energy efficiency gains; and on Hurricane, \abbr~achieves $4.90\times$--$37.04\times$ runtime speedups alongside $35.02\times$--$200.78\times$ energy efficiency improvements. These consistent gains across diverse datasets and platforms highlight the generality and scalability of our architecture for both compression and decompression processes.

\subsection{Ablation Study}

\begin{figure}[t]
  \centering
  \begin{subfigure}{\linewidth}
    \centering
    \includegraphics[width=0.9\linewidth]{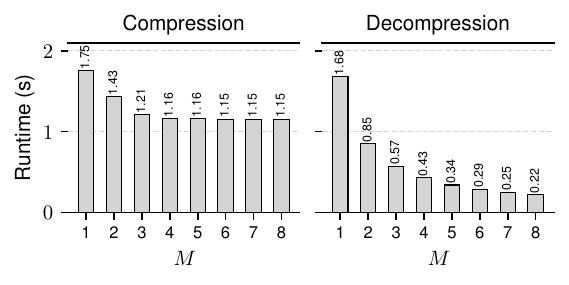}
    \vspace{-3mm}
    \caption{Data-size scalability. Compression and decompression runtime on the Nyx dataset with $M$ scaled from 1 to 8.}
    \label{fig:scalability_M}
  \end{subfigure}
  \begin{subfigure}{\linewidth}
    \centering
    \includegraphics[width=0.9\linewidth]{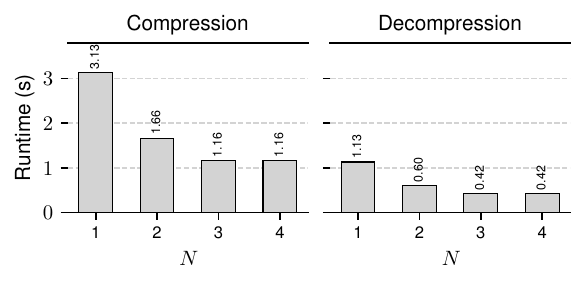}
    \vspace{-3mm}
    \caption{Workload scalability. Compression and decompression runtime for two Nyx, one Miranda, and one Hurricane datasets with $N$ scaled from 1 to 4.}
    \label{fig:scalability_N}
  \end{subfigure}
  \vspace{-1\baselineskip}
  \caption{Scalability analysis of the proposed \abbr.}
  \vspace{-4.5mm}
  \label{fig:scalability}
\end{figure}

\textbf{Scalability of the architecture.} The modular architecture is designed with scalability as a core principle, offering two dimensions of flexibility: data size scalability and workload scalability.

For data-size scalability, the architecture adjusts the number of 1D systolic arrays ($M$) within each \abbr~Computing Core to support different input sizes. Increasing $M$ allows a single core to process more data in parallel and reduce runtime. As shown in Figure~\ref{fig:scalability}, we evaluate the compression and decompression runtime of the Nyx dataset by scaling $M$ from 1 to 8. Decompression time decreases clearly as $M$ increases. However, compression speedup is less pronounced because the fully pipelined dataflow and neural network training make the Neural Engine the dominant bottleneck. For example, the runtime reaches its minimum at $M=4$, and adding more systolic arrays brings little benefit.

For workload scalability, the architecture supports parallel execution across multiple \abbr~Computing Cores ($N$), each independently processing a distinct dataset or task. This is useful for large-scale deployments where multiple workloads must run concurrently. We evaluate runtime while processing two Nyx datasets, one Miranda dataset, and one Hurricane dataset as $N$ scales from 1 to 4. As shown in Figure~\ref{fig:scalability}, runtime consistently decreases with more cores. When scaling from $N=3$ to $N=4$, the two Nyx datasets become the dominant bottleneck, with runtime reaching 1.16 seconds for compression and 0.42 seconds for decompression.

These results demonstrate that the proposed architecture effectively scales along both data size and workload dimensions. The overall speedup depends on workload distribution and system bottlenecks—particularly those introduced by the neural network or dataset-specific characteristics~\cite{liu2025performance}.

\textbf{Data movement reduction.} To quantify the impact of the proposed architecture, we analyze internal data transfers within a single \abbr~Computing Core, focusing on the Prediction, Neural, and Codec Engines as the main contributors during compression and decompression.\footnote{The data access during neural network training and inference is optimized via TimeLoop and excluded from this analysis. Thus, the data access pattern is identical in both compression and decompression procedures.} In the Prediction Engine, look-ahead processing reuses intermediate results from on-chip SRAM, while the systolic arrays keep data local within processing elements. Across engines, the hybrid dataflow supports pipelined and parallel execution, reducing unnecessary off-chip communication. Operator fusion further removes separate memory accesses for slice-wise normalization and convolution. Together, these features reduce overall data movement by improving on-chip reuse.

\begin{figure}[t]
  \centering
  \includegraphics[width=\linewidth]{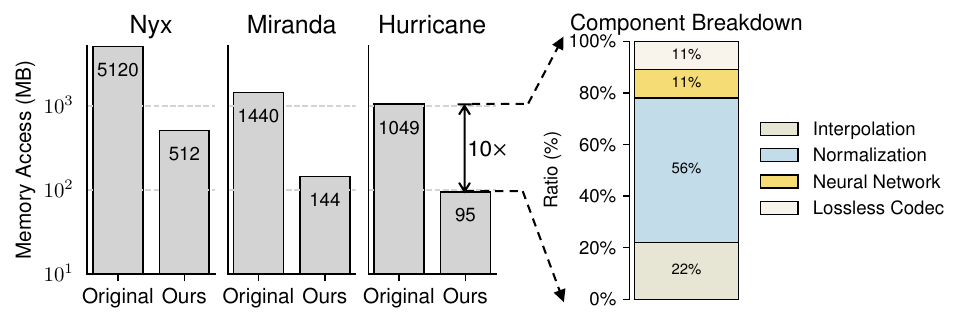}
  \vspace{-7mm}
  \caption{Off-chip data movement reduction of \abbr.}
  \label{fig:dataMovement}
\end{figure}

As shown in Figure~\ref{fig:dataMovement}, the proposed architecture reduces data movement by up to $10\times$ across three datasets. Normalization contributes the largest share of the reduction (56\%), followed by prediction (22\%), neural network operations (11\%), and lossless codec (11\%). These results show the proposed design effectively reduces data movement across compression and decompression workloads.

\section{Conclusion}

This paper presents \abbr, a dataflow-aware and scalable hardware architecture for accelerating neural-hybrid scientific lossy compression. \abbr~uses look-ahead computation to reduce SRAM usage and improve pipeline throughput, and fuses slice-wise normalization into convolution to reduce bubble overhead and data movement. The proposed \abbr~Computing Core improves on-chip data reuse, supports efficient hybrid execution, and scales with both dataset size and workload volume. To our knowledge, this is the first customized design for scientific lossy compression that jointly addresses dataflow efficiency and architectural scalability. Experiments show clear latency and energy-efficiency improvements over state-of-the-art baselines.

\section*{Acknowledgment}
This work was supported in part by the U.S. National Science Foundation CAREER Award No. 2544246, by the U.S. National Science Foundation under Grants OAC-2311875, OAC-2514036, OAC-2514034, OAC-2609480, OAC-2513768, SHF-2331301, and SHF-2508118, and by the U.S. Department of Energy, Office of Science, Advanced Scientific Computing Research (ASCR), under Contract DE-AC02-06CH11357. The research described herein was also funded by the Generative AI for Science, Energy, and Security Science \& Technology Investment under the Laboratory Directed Research and Development Program at Pacific Northwest National Laboratory (PNNL). PNNL is a multiprogram national laboratory operated for the U.S. Department of Energy (DOE) by Battelle Memorial Institute under Contract No. DE-AC05-76RL0-1830.

\FloatBarrier

\bibliographystyle{acm-ref-format}
\bibliography{refs}

\appendix

\end{document}

%% file: preamble.tex
\definecolor{miao}{RGB}{140,114,143}

\usepackage{hyperref}
\hypersetup{colorlinks=true,linkcolor=blue,urlcolor=blue,citecolor=miao}

\usepackage{listings}

\lstdefinestyle{cuda}{
	belowcaptionskip=1\baselineskip,
	breaklines=true,
	xleftmargin=\parindent,
	language=C,  
	showstringspaces=false,
	basicstyle=\footnotesize\ttfamily,
	keywordstyle=\bfseries\color{PineGreen},
	commentstyle=\color{white!40!black},
	identifierstyle=\color{NavyBlue},
	stringstyle=\color{orange},
	escapeinside={(*@}{@*)},
	morekeywords={shared, is, to},
	numbers=left,
	morekeywords={[2]{copy,syncthreads,nnz}},
	keywordstyle={[2]{\color{Purple}}},
}     
\def\BibTeX{{\rm B\kern-.05em{\sc i\kern-.025em b}\kern-.08em
    T\kern-.1667em\lower.7ex\hbox{E}\kern-.125emX}}

\usepackage{tikz}
\newcommand*\Circled[2][gray!40]{%
	\tikz[baseline=(char.base)]{\node[
        shape=circle, draw=none,  thick, 
        fill=#1 ,inner sep=0.9pt] (char) 
    {\textcolor{black}{#2}}; 
}}

\usepackage{framed}

\definecolor{that-yellow}{HTML}{ddc26f}

\makeatletter
\let\@authorsaddresses\@empty
\makeatother

%% file: refs.bib
@inproceedings{CPU-HBM-not-working,
  author    = {McCalpin, John D.},
  editor    = {Bienz, Amanda
               and Weiland, Mich{\`e}le
               and Baboulin, Marc
               and Kruse, Carola},
  title     = {Bandwidth Limits in the Intel Xeon Max (Sapphire Rapids with HBM) Processors},
  booktitle = {High Performance Computing},
  year      = {2023},
  publisher = {Springer Nature Switzerland},
  address   = {Cham},
  pages     = {403--413},
  isbn      = {978-3-031-40843-4}
}

@article{cappello2025lossy,
  title={Lossy Compression of Scientific Data: Applications Constrains and Requirements},
  author={Cappello, Franck and Baker, Allison and Bozda, Ebru and Burtscher, Martin and Chard, Kyle and Di, Sheng and Grady, Paul Christopher O and Jiang, Peng and Li, Shaomeng and Lindahl, Erik and others},
  journal={arXiv preprint arXiv:2503.20031},
  year={2025}
}

@article{rosa2020data,
  title={Data Science Strategies for Multimessenger Astronomy},
  author={Rosa, Reinaldo R},
  journal={Anais da Academia Brasileira de Ci{\^e}ncias},
  volume={93},
  pages={e20200861},
  year={2020},
  publisher={SciELO Brasil}
}

@article{di2024survey,
  title={A survey on error-bounded lossy compression for scientific datasets},
  author={Di, Sheng and Liu, Jinyang and Zhao, Kai and Liang, Xin and Underwood, Robert and Zhang, Zhaorui and Shah, Milan and Huang, Yafan and Huang, Jiajun and Yu, Xiaodong and others},
  journal={arXiv preprint arXiv:2404.02840},
  year={2024}
}

@inproceedings{liu2024cusz,
  title={CUSZ-i: High-Ratio Scientific Lossy Compression on GPUs with Optimized Multi-Level Interpolation},
  author={Liu, Jinyang and Tian, Jiannan and Wu, Shixun and Di, Sheng and Zhang, Boyuan and Underwood, Robert and Huang, Yafan and Huang, Jiajun and Zhao, Kai and Li, Guanpeng and others},
  booktitle={SC24: International Conference for High Performance Computing, Networking, Storage and Analysis},
  pages={1--15},
  year={2024},
  organization={IEEE}
}

@inproceedings{tian2020cusz,
  title={Cusz: An efficient gpu-based error-bounded lossy compression framework for scientific data},
  author={Tian, Jiannan and Di, Sheng and Zhao, Kai and Rivera, Cody and Fulp, Megan Hickman and Underwood, Robert and Jin, Sian and Liang, Xin and Calhoun, Jon and Tao, Dingwen and others},
  booktitle={Proceedings of the ACM International Conference on Parallel Architectures and Compilation Techniques},
  pages={3--15},
  year={2020}
}

@inproceedings{huang2024cuszp2,
  title={cuSZp2: A GPU Lossy Compressor with Extreme Throughput and Optimized Compression Ratio},
  author={Huang, Yafan and Di, Sheng and Li, Guanpeng and Cappello, Franck},
  booktitle={SC24: International Conference for High Performance Computing, Networking, Storage and Analysis},
  pages={1--18},
  year={2024},
  organization={IEEE}
}

@article{meehl2000coupled,
  title={The coupled model intercomparison project (CMIP)},
  author={Meehl, Gerald A and Boer, George J and Covey, Curt and Latif, Mojib and Stouffer, Ronald J},
  journal={Bulletin of the American Meteorological Society},
  volume={81},
  number={2},
  pages={313--318},
  year={2000},
  publisher={JSTOR}
}

@article{johnson2023exploring,
author = {Johnson, Nnadikwe},
year = {2023},
month = {12},
pages = {},
title = {Exploring Cutting-Edge Application of Computational Fluid Dynamic in Enhancing and Innovation within the Oil and Gas Sector},
doi = {10.20944/preprints202312.1113.v1}
}

@article{ziv1977universal,
  title     = {A universal algorithm for sequential data compression},
  author    = {Ziv, Jacob and Lempel, Abraham},
  journal   = {IEEE Transactions on information theory},
  volume    = {23},
  number    = {3},
  pages     = {337--343},
  year      = {1977},
  publisher = {IEEE}
}

@misc{GZIP1996,
  author    = {Deutsch, P.},
  title     = {RFC1952: GZIP file format specification version 4.3},
  year      = {1996},
  publisher = {RFC Editor},
  address   = {USA}
}

@article{lindstrom2006fast,
  title     = {Fast and efficient compression of floating-point data},
  author    = {Lindstrom, Peter and Isenburg, Martin},
  journal   = {IEEE transactions on visualization and computer graphics},
  volume    = {12},
  number    = {5},
  pages     = {1245--1250},
  year      = {2006},
  publisher = {IEEE}
}

@misc{gailly2004zlib,
  author      = {Gailly, Jean-loup and Adler, Mark},
  title       = {zlib Compression Library},
  year        = {2004},
  type        = {Other},
  institution = {Apollo - University of Cambridge Repository},
  url         = {http://www.dspace.cam.ac.uk/handle/1810/3486},
  note        = {Accessed: 2025-01-08}
}

@inproceedings{zhao2020sdrbench,
  author    = { Zhao, Kai and Di, Sheng and Lian, Xin and Li, Sihuan and Tao, Dingwen and Bessac, Julie and Chen, Zizhong and Cappello, Franck },
  booktitle = { 2020 IEEE International Conference on Big Data (Big Data) },
  title     = {{ SDRBench: Scientific Data Reduction Benchmark for Lossy Compressors }},
  year      = {2020},
  volume    = {},
  issn      = {},
  pages     = {2716-2724},
  doi       = {10.1109/BigData50022.2020.9378449},
  url       = {https://doi.ieeecomputersociety.org/10.1109/BigData50022.2020.9378449},
  publisher = {IEEE Computer Society},
  address   = {Los Alamitos, CA, USA},
  month     = Dec
}

@article{lindstrom2014fixed,
  title     = {Fixed-rate compressed floating-point arrays},
  author    = {Lindstrom, Peter},
  journal   = {IEEE transactions on visualization and computer graphics},
  volume    = {20},
  number    = {12},
  pages     = {2674--2683},
  year      = {2014},
  publisher = {IEEE}
}

@inproceedings{di2016sz,
  author    = { Di, Sheng and Cappello, Franck },
  booktitle = { 2016 IEEE International Parallel and Distributed Processing Symposium (IPDPS) },
  title     = {{ Fast Error-Bounded Lossy HPC Data Compression with SZ }},
  year      = {2016},
  volume    = {},
  issn      = {1530-2075},
  pages     = {730-739},
  doi       = {10.1109/IPDPS.2016.11},
  url       = {https://doi.ieeecomputersociety.org/10.1109/IPDPS.2016.11},
  publisher = {IEEE Computer Society},
  address   = {Los Alamitos, CA, USA},
  month     = May
}

@inproceedings{liu2022qoz,
  author    = { Liu, Jinyang and Di, Sheng and Zhao, Kai and Liang, Xin and Chen, Zizhong and Cappello, Franck },
  booktitle = { SC22: International Conference for High Performance Computing, Networking, Storage and Analysis },
  title     = {{ Dynamic Quality Metric Oriented Error Bounded Lossy Compression for Scientific Datasets }},
  year      = {2022},
  volume    = {},
  issn      = {},
  pages     = {1-15},
  doi       = {10.1109/SC41404.2022.00067},
  url       = {https://doi.ieeecomputersociety.org/10.1109/SC41404.2022.00067},
  publisher = {IEEE Computer Society},
  address   = {Los Alamitos, CA, USA},
  month     = Nov
}

@inproceedings{zhao2021splinesz,
  author    = { Zhao, Kai and Di, Sheng and Dmitriev, Maxim and Tonellot, Thierry-Laurent D. and Chen, Zizhong and Cappello, Franck },
  booktitle = { 2021 IEEE 37th International Conference on Data Engineering (ICDE) },
  title     = {{ Optimizing Error-Bounded Lossy Compression for Scientific Data by Dynamic Spline Interpolation }},
  year      = {2021},
  volume    = {},
  issn      = {},
  pages     = {1643-1654},
  doi       = {10.1109/ICDE51399.2021.00145},
  url       = {https://doi.ieeecomputersociety.org/10.1109/ICDE51399.2021.00145},
  publisher = {IEEE Computer Society},
  address   = {Los Alamitos, CA, USA},
  month     = apr
}

@inproceedings{tao2017significant,
  author    = { Tao, Dingwen and Di, Sheng and Chen, Zizhong and Cappello, Franck },
  booktitle = { 2017 IEEE International Parallel and Distributed Processing Symposium (IPDPS) },
  title     = {{ Significantly Improving Lossy Compression for Scientific Data Sets Based on Multidimensional Prediction and Error-Controlled Quantization }},
  year      = {2017},
  volume    = {},
  issn      = {1530-2075},
  pages     = {1129-1139},
  doi       = {10.1109/IPDPS.2017.115},
  url       = {https://doi.ieeecomputersociety.org/10.1109/IPDPS.2017.115},
  publisher = {IEEE Computer Society},
  address   = {Los Alamitos, CA, USA},
  month     = Jun
}

@inproceedings{liang2019hybrid,
  author    = {Liang, Xin and Di, Sheng and Li, Sihuan and Tao, Dingwen and Nicolae, Bogdan and Chen, Zizhong and Cappello, Franck},
  title     = {Significantly improving lossy compression quality based on an optimized hybrid prediction model},
  year      = {2019},
  isbn      = {9781450362290},
  publisher = {Association for Computing Machinery},
  address   = {New York, NY, USA},
  url       = {https://doi.org/10.1145/3295500.3356193},
  doi       = {10.1145/3295500.3356193},
  booktitle = {Proceedings of the International Conference for High Performance Computing, Networking, Storage and Analysis},
  articleno = {33},
  numpages  = {26},
  location  = {Denver, Colorado},
  series    = {SC '19}
}

@article{tao2019optimizing,
  title     = {Optimizing lossy compression rate-distortion from automatic online selection between SZ and ZFP},
  author    = {Tao, Dingwen and Di, Sheng and Liang, Xin and Chen, Zizhong and Cappello, Franck},
  journal   = {IEEE Transactions on Parallel and Distributed Systems},
  volume    = {30},
  number    = {8},
  pages     = {1857--1871},
  year      = {2019},
  publisher = {IEEE}
}

@article{liang2022sz3,
  title={Sz3: A modular framework for composing prediction-based error-bounded lossy compressors},
  author={Liang, Xin and Zhao, Kai and Di, Sheng and Li, Sihuan and Underwood, Robert and Gok, Ali M and Tian, Jiannan and Deng, Junjing and Calhoun, Jon C and Tao, Dingwen and others},
  journal={IEEE Transactions on Big Data},
  volume={9},
  number={2},
  pages={485--498},
  year={2022},
  publisher={IEEE}
}

@article{hornik1989multilayer,
  title     = {Multilayer feedforward networks are universal approximators},
  author    = {Hornik, Kurt and Stinchcombe, Maxwell and White, Halbert},
  journal   = {Neural networks},
  volume    = {2},
  number    = {5},
  pages     = {359--366},
  year      = {1989},
  publisher = {Elsevier}
}

@inproceedings{liu2021autoencoder,
  title={Exploring autoencoder-based error-bounded compression for scientific data},
  author={Liu, Jinyang and Di, Sheng and Zhao, Kai and Jin, Sian and Tao, Dingwen and Liang, Xin and Chen, Zizhong and Cappello, Franck},
  booktitle={2021 IEEE International Conference on Cluster Computing (CLUSTER)},
  pages={294--306},
  year={2021},
  organization={IEEE}
}

@inproceedings{liu2023srnsz,
  title={Scientific error-bounded lossy compression with super-resolution neural networks},
  author={Liu, Jinyang and Di, Sheng and Jin, Sian and Zhao, Kai and Liang, Xin and Chen, Zizhong and Cappello, Franck},
  booktitle={2023 IEEE International Conference on Big Data (BigData)},
  pages={229--236},
  year={2023},
  organization={IEEE}
}

@inproceedings{liu2025crossfield,
  title     = {Advancing Scientific Data Compression via Cross-Field Prediction},
  author    = {Liu, Youyuan and Jia, Wenqi and Yang, Taolue and Jiang, Bo and Yin, Miao and Jin, Sian},
  booktitle = {Proceedings of the 34th International Symposium on High-Performance Parallel and Distributed Computing},
  pages     = {1--12},
  year      = {2025},
  publisher = {ACM},
  doi       = {10.1145/3731545.3731592}
}

@misc{li2024attention,
  title         = {Attention Based Machine Learning Methods for Data Reduction with Guaranteed Error Bounds},
  author        = {Xiao Li and Jaemoon Lee and Anand Rangarajan and Sanjay Ranka},
  year          = {2024},
  eprint        = {2409.05357},
  archiveprefix = {arXiv},
  primaryclass  = {cs.LG},
  url           = {https://arxiv.org/abs/2409.05357}
}

@inproceedings{gwlz,
  author    = {Jia, Wenqi and Jin, Sian and Wang, Jinzhen and Niu, Wei and Tao, Dingwen and Yin, Miao},
  title     = {GWLZ: A Group-wise Learning-based Lossy Compression Framework for Scientific Data},
  year      = {2024},
  isbn      = {9798400706424},
  publisher = {Association for Computing Machinery},
  address   = {New York, NY, USA},
  url       = {https://doi.org/10.1145/3659995.3660041},
  doi       = {10.1145/3659995.3660041},
  booktitle = {Proceedings of the 14th Workshop on AI and Scientific Computing at Scale Using Flexible Computing Infrastructures},
  pages     = {34–41},
  numpages  = {8},
  location  = {Pisa, Italy},
  series    = {FlexScience'24}
}

@inproceedings{jia2024neurlz,
  title     = {{NeurLZ}: An Online Neural Learning-Based Method to Enhance Scientific Lossy Compression},
  author    = {Jia, Wenqi and Hu, Zhewen and Liu, Youyuan and Zhang, Boyuan and Wang, Jinzhen and Liu, Jinyang and Niu, Wei and Kalafatis, Stavros and Huang, Junzhou and Jin, Sian and Wang, Daoce and Tian, Jiannan and Yin, Miao},
  booktitle = {Proceedings of the 39th ACM International Conference on Supercomputing},
  pages     = {26--42},
  year      = {2025},
  publisher = {ACM},
  doi       = {10.1145/3721145.3725763}
}

@inproceedings{wang2025stz,
  title     = {{STZ}: A High Quality and High Speed Streaming Lossy Compression Framework for Scientific Data},
  author    = {Wang, Daoce and Grosset, Pascal and Pulido, Jesus and Tian, Jiannan and Athawale, Tushar and Jia, Jinda and Sun, Baixi and Zhang, Boyuan and Jin, Sian and Zhao, Kai and Ahrens, James and Song, Fengguang},
  booktitle = {Proceedings of the International Conference for High Performance Computing, Networking, Storage and Analysis},
  pages     = {2038--2055},
  year      = {2025},
  publisher = {ACM},
  doi       = {10.1145/3712285.3759795}
}

@inproceedings{jiao2025adaptive,
  title     = {Improving the Efficiency of Interpolation-Based Scientific Data Compressors with Adaptive Quantization Index Prediction},
  author    = {Jiao, Pu and Di, Sheng and Xia, Mingze and Wu, Xuan and Liu, Jinyang and Liang, Xin and Cappello, Franck},
  booktitle = {2025 IEEE International Parallel and Distributed Processing Symposium},
  pages     = {974--986},
  year      = {2025},
  publisher = {IEEE},
  doi       = {10.1109/IPDPS64566.2025.00091}
}

@inproceedings{wu2025unstructured,
  title     = {Enabling Efficient Error-Controlled Lossy Compression for Unstructured Scientific Data},
  author    = {Wu, Xuan and Di, Sheng and Ren, Congrong and Jiao, Pu and Xia, Mingze and Wang, Cheng and Guo, Hanqi and Liang, Xin and Cappello, Franck},
  booktitle = {2025 IEEE International Parallel and Distributed Processing Symposium},
  pages     = {370--382},
  year      = {2025},
  publisher = {IEEE},
  doi       = {10.1109/IPDPS64566.2025.00040}
}

@inproceedings{yang2025ipcomp,
  title     = {{IPComp}: Interpolation Based Progressive Lossy Compression for Scientific Applications},
  author    = {Yang, Zhuoxun and Di, Sheng and Zhang, Longtao and Li, Ruoyu and Li, Ximiao and Huang, Jiajun and Liu, Jinyang and Cappello, Franck and Zhao, Kai},
  booktitle = {Proceedings of the 34th International Symposium on High-Performance Parallel and Distributed Computing},
  pages     = {1--14},
  year      = {2025},
  publisher = {ACM},
  doi       = {10.1145/3731545.3731578}
}

@article{zhang2025lcp,
  title     = {{LCP}: Enhancing Scientific Data Management with Lossy Compression for Particles},
  author    = {Zhang, Longtao and Li, Ruoyu and Ren, Congrong and Di, Sheng and Liu, Jinyang and Huang, Jiajun and Underwood, Robert and Grosset, Pascal and Tao, Dingwen and Liang, Xin and Guo, Hanqi and Cappello, Franck and Zhao, Kai},
  journal   = {Proceedings of the ACM on Management of Data},
  volume    = {3},
  number    = {1},
  pages     = {1--27},
  year      = {2025},
  publisher = {ACM},
  doi       = {10.1145/3709700}
}

@inproceedings{zhang2026opal,
  title     = {{OPAL}: On-Demand Progressive Accelerated Scientific Lossy Compression},
  author    = {Zhang, Longtao and Li, Ruoyu and Yang, Zhuoxun and Underwood, Robert and Di, Sheng and Wang, Daoce and Liu, Jinyang and Huang, Jiajun and Cappello, Franck and Zhao, Kai},
  booktitle = {Proceedings of the 35th International Symposium on High-Performance Parallel and Distributed Computing},
  pages     = {58--70},
  year      = {2026},
  publisher = {ACM},
  doi       = {10.1145/3806645.3807577}
}

@article{yan2025eureka,
  title     = {{Eureka}: Enabling Fine-Grained Access and Range Queries on Compressed Scientific Data via Data-Index Co-Compression},
  author    = {Yan, Ning and Di, Sheng and Zhao, Kai and Wan, Lipeng},
  journal   = {Proceedings of the VLDB Endowment},
  volume    = {19},
  number    = {4},
  pages     = {631--643},
  year      = {2025},
  publisher = {VLDB Endowment},
  doi       = {10.14778/3785297.3785305}
}

@inproceedings{xia2024determinant,
  title     = {Preserving Topological Feature with Sign-of-Determinant Predicates in Lossy Compression: A Case Study of Vector Field Critical Points},
  author    = {Xia, Mingze and Di, Sheng and Cappello, Franck and Jiao, Pu and Zhao, Kai and Liu, Jinyang and Wu, Xuan and Liang, Xin and Guo, Hanqi},
  booktitle = {2024 IEEE 40th International Conference on Data Engineering (ICDE)},
  pages     = {4979--4992},
  year      = {2024},
  publisher = {IEEE},
  doi       = {10.1109/ICDE60146.2024.00378}
}

@inproceedings{xia2025tspsz,
  title     = {{TspSZ}: An Efficient Parallel Error-Bounded Lossy Compressor for Topological Skeleton Preservation},
  author    = {Xia, Mingze and Wang, Bei and Li, Yuxiao and Jiao, Pu and Liang, Xin and Guo, Hanqi},
  booktitle = {2025 IEEE 41st International Conference on Data Engineering (ICDE)},
  pages     = {3682--3695},
  year      = {2025},
  publisher = {IEEE},
  doi       = {10.1109/ICDE65448.2025.00275}
}

@inproceedings{liu2025performance,
  title     = {Accurate Performance Modeling and Uncertainty Analysis of Lossy Compression in Scientific Applications},
  author    = {Liu, Youyuan and Yang, Taolue and Jin, Sian},
  booktitle = {2025 Data Compression Conference (DCC)},
  pages     = {390--390},
  year      = {2025},
  publisher = {IEEE},
  doi       = {10.1109/DCC62719.2025.00077}
}

@inproceedings{yang2024seqbench,
  title     = {{SeqBench}: A Benchmark Suite for Lossless and Lossy Compression of Sequence Data},
  author    = {Yang, Taolue and Liu, Youyuan and Li, Chong and Shi, Xinhua Mindy and Jin, Sian},
  booktitle = {Proceedings of the 15th ACM International Conference on Bioinformatics, Computational Biology and Health Informatics},
  pages     = {1--7},
  year      = {2024},
  publisher = {ACM},
  doi       = {10.1145/3698587.3701386}
}

@inproceedings{xia2026trajectories,
  title     = {Time-Varying Vector Field Compression with Preserved Critical Point Trajectories},
  author    = {Xia, Mingze and Li, Yuxiao and Jiao, Pu and Wang, Bei and Liang, Xin and Guo, Hanqi},
  booktitle = {2026 IEEE 42nd International Conference on Data Engineering (ICDE)},
  pages     = {2235--2249},
  year      = {2026},
  publisher = {IEEE},
  doi       = {10.1109/ICDE65706.2026.00168}
}

@article{lakshminarasimhan2013isabela,
  title     = {ISABELA for effective in situ compression of scientific data},
  author    = {Lakshminarasimhan, Sriram and Shah, Neil and Ethier, Stephane and Ku, Seung-Hoe and Chang, Choong-Seock and Klasky, Scott and Latham, Rob and Ross, Rob and Samatova, Nagiza F},
  journal   = {Concurrency and Computation: Practice and Experience},
  volume    = {25},
  number    = {4},
  pages     = {524--540},
  year      = {2013},
  publisher = {Wiley Online Library}
}

@article{almgren2013nyx,
  title     = {Nyx: A massively parallel amr code for computational cosmology},
  author    = {Almgren, Ann S and Bell, John B and Lijewski, Mike J and Luki{\'c}, Zarija and Van Andel, Ethan},
  journal   = {The Astrophysical Journal},
  volume    = {765},
  number    = {1},
  pages     = {39},
  year      = {2013},
  publisher = {IOP Publishing}
}

@article{sexton2021nyx,
  title={Nyx: A massively parallel amr code for computational cosmology},
  author={Sexton, Jean and Lukic, Zarija and Almgren, Ann and Daley, Chris and Friesen, Brian and Myers, Andrew and Zhang, Weiqun},
  journal={The Journal of Open Source Software},
  volume={6},
  number={63},
  pages={3068},
  year={2021}
}

@article{lukic2015lyman,
  title     = {The Lyman $\alpha$ forest in optically thin hydrodynamical simulations},
  author    = {Luki{\'c}, Zarija and Stark, Casey W and Nugent, Peter and White, Martin and Meiksin, Avery A and Almgren, Ann},
  journal   = {Monthly Notices of the Royal Astronomical Society},
  volume    = {446},
  number    = {4},
  pages     = {3697--3724},
  year      = {2015},
  publisher = {Oxford University Press}
}

@article{onorbe2019inhomogeneous,
  title     = {Inhomogeneous reionization models in cosmological hydrodynamical simulations},
  author    = {O{\~n}orbe, Jose and Davies, FB and Luki{\'c}, Z and Hennawi, JF and Sorini, D},
  journal   = {Monthly Notices of the Royal Astronomical Society},
  volume    = {486},
  number    = {3},
  pages     = {4075--4097},
  year      = {2019},
  publisher = {Oxford University Press}
}

@article{kim2023design,
  title={Design and implementation of I/O performance prediction scheme on HPC systems through large-scale log analysis},
  author={Kim, Sunggon and Sim, Alex and Wu, Kesheng and Byna, Suren and Son, Yongseok},
  journal={Journal of Big Data},
  volume={10},
  number={1},
  pages={65},
  year={2023},
  publisher={Springer}
}

@article{behzad2019optimizing,
  title={Optimizing i/o performance of hpc applications with autotuning},
  author={Behzad, Babak and Byna, Surendra and Prabhat and Snir, Marc},
  journal={ACM Transactions on Parallel Computing (TOPC)},
  volume={5},
  number={4},
  pages={1--27},
  year={2019},
  publisher={ACM New York, NY, USA}
}

@inproceedings{paul2020understanding,
  title={Understanding hpc application i/o behavior using system level statistics},
  author={Paul, Arnab K and Faaland, Olaf and Moody, Adam and Gonsiorowski, Elsa and Mohror, Kathryn and Butt, Ali R},
  booktitle={2020 IEEE 27th International Conference on High Performance Computing, Data, and Analytics (HiPC)},
  pages={202--211},
  year={2020},
  organization={IEEE}
}

@inproceedings{chunduri2019gpcnet,
  title={Gpcnet: Designing a benchmark suite for inducing and measuring contention in hpc networks},
  author={Chunduri, Sudheer and Groves, Taylor and Mendygral, Peter and Austin, Brian and Balma, Jacob and Kandalla, Krishna and Kumaran, Kalyan and Lockwood, Glenn and Parker, Scott and Warren, Steven and others},
  booktitle={Proceedings of the International Conference for High Performance Computing, Networking, Storage and Analysis},
  pages={1--33},
  year={2019}
}

@inproceedings{kang2022study,
  title={Study of workload interference with intelligent routing on dragonfly},
  author={Kang, Yao and Wang, Xin and Lan, Zhiling},
  booktitle={SC22: International Conference for High Performance Computing, Networking, Storage and Analysis},
  pages={1--14},
  year={2022},
  organization={IEEE}
}

@inproceedings{parashar2019timeloop,
  title={Timeloop: A systematic approach to dnn accelerator evaluation},
  author={Parashar, Angshuman and Raina, Priyanka and Shao, Yakun Sophia and Chen, Yu-Hsin and Ying, Victor A and Mukkara, Anurag and Venkatesan, Rangharajan and Khailany, Brucek and Keckler, Stephen W and Emer, Joel},
  booktitle={2019 IEEE international symposium on performance analysis of systems and software (ISPASS)},
  pages={304--315},
  year={2019},
  organization={IEEE}
}
